\documentclass{jpsj-suppl}
\usepackage{txfonts} 
\usepackage{graphicx,amsmath,wrapfig,epsfig}

\title{Exotic-Hadron Signature by Constituent-Counting Rule \\ in Perturbative QCD}
\author{
Wen-Chen \textsc{Chang}$^1$
H. \textsc{Kawamura}$^2$, S. \textsc{Kumano}$^{3,4}$, 
and T. \textsc{Sekihara}$^5$
}
\inst{
$^1$ 
Institute of Physics, Academia Sinica, Taipei 11529, Taiwan \\
$^2$ 
Department of Mathematics, Juntendo University, Inzai, Chiba,
270-1695, Japan \\
$^3$KEK Theory Center, Institute of Particle and Nuclear Studies, KEK, \\
\ \ 1-1, Ooho, Tsukuba, Ibaraki, 305-0801, Japan \\
$^4$J-PARC Branch, KEK Theory Center,
Institute of Particle and Nuclear Studies, KEK \\
\ \ and Theory Group, Particle and Nuclear Physics Division, 
J-PARC Center, \\
\ \ 203-1, Shirakata, Tokai, Ibaraki, 319-1106, Japan \\
$^5$ Advanced Science Research Center, Japan Atomic Energy Agency, \\
\ \ Shirakata, Tokai, Ibaraki, 319-1195, Japan
}
\email{shunzo.kumano@kek.jp}

\recdate{October 17, 2016}

\abst{
We explain a method to find internal quark configurations of
exotic hadron candidates by using the constituent counting rule.
The counting rule was theoretically predicted 
in perturbative QCD for hard exclusive hadron reactions,
and it has been tested in experiments for stable hadrons
including compound systems of hadrons such as the deuteron, 
$^3$H, and $^3$He. It indicates that the cross section scales
as $d\sigma /dt \sim 1/s^{n-2}$, where $s$ is the center-of-mass
energy squared and $n$ is the total number of constituents.
We apply this method for finding internal
configurations of exotic hadron candidates, especially 
$\Lambda (1405)$. There is a possibility that $\Lambda (1405)$
could be five-quark state or a $\bar K N$ molecule, and scaling
properties should be different between the ordinary three-quark state
or five-quark one. We predict such a difference in 
$\pi^- + p \to K^0 + \Lambda (1405)$, and it could be experimentally
tested, for example, at J-PARC. On the other hand, there are already
measurements for $\gamma + p \to K^+ + \Lambda (1405)$ as well as
the ground $\Lambda$ in photoproduction reactions. Analyzing 
such data, we found an interesting indication that $\Lambda (1405)$
looks like a five-quark state at medium energies and a three-quark one
at high energies. However, accurate higher-energy measurements are 
necessary for drawing a solid conclusion, and it should be done at 
JLab by using the updated 12 GeV electron beam. Furthermore, we discuss
studies of exotic hadron candidates, such as $f_0 (980)$ and 
$a_0 (980)$, in electron-positron annihilation by using
generalized distribution amplitudes and the counting rule.
These studies should be possible as a KEKB experiment.
}

\kword{hadron, exotic, exclusive, QCD, quark, gluon}

\begin{document}
\maketitle

\section{Introduction}

Many observed hadrons are interpreted by the ordinary quark compositions,
$q\bar q$ and $qqq$, according to constituent quark models. Since
other quark configurations are not prohibited by QCD, they have been
searched for a long time. They are called exotic hadrons, and 
typical ones are tetraquark ($qq\bar q\bar q$) and pentaquark
($qqqq\bar q$) hadrons. There are recent experimental findings
about candidates of these exotic hadrons \cite{pdg-2016}.
However, it is difficult to find an undoubted evidence of 
exotic nature by global observables such as masses, 
spins, parities, and decay widths.
Since appropriate degrees of freedom are quarks and gluons at high energies,
there is a possibility that the exotic evidence is found in high-energy
reactions. Along this line, we have proposed exotic-hadron studies
by using fragmentation functions \cite{ffs-exotic}, constituent
counting rule \cite{kks2013,cks2016}, and three-dimensional structure 
functions such as generalized parton distributions (GPDs)
and generalized distribution amplitudes (GDAs) \cite{gpd-gda}.
There are also other recent studies on using the counting rule
for exotic hadrons \cite{counting-other}.

Among these studies, we explain our works on the constituent 
counting rule for exotic hadron studies \cite{kks2013,cks2016,gpd-gda}.
The constituent counting rule is theoretically predicated
in perturbative QCD for exclusive reactions with large transverse
momentum \cite{exclusive-theory}, and it has been confirmed by
hadron-hadron and lepton-hadron reactions \cite{counting-exp}
including light nuclei \cite{counting-exp-nuclei}.
An exclusive cross section scales as $1/s^{n-2}$ with the total
number of constituents $n$ and the center-of-mass energy squared $s$,
so that it should be used for exotic hadron studies
\cite{kks2013,cks2016,gpd-gda,counting-other}.
In this article, we explain the constituent counting rule in 
Sec.\,\ref{counting}, and then it is used for the exotic hadron 
candidate $\Lambda (1405)$ to compare with experimental data
and to propose new experiments in Sec.\,\ref{lambda-1405}.
Next, possible exotic hadron studies are discussed in Sec.\,\ref{3d}
by using the GPDs and GDAs. We summarize our studies in Sec.\,\ref{summary}.

\section{Constituent-Counting Rule in Hard Exclusive Hadron Production}
\label{counting}

\begin{wrapfigure}[10]{r}{0.40\textwidth}
   \vspace{-0.15cm}
   \begin{center}
     \includegraphics[width=3.5cm]{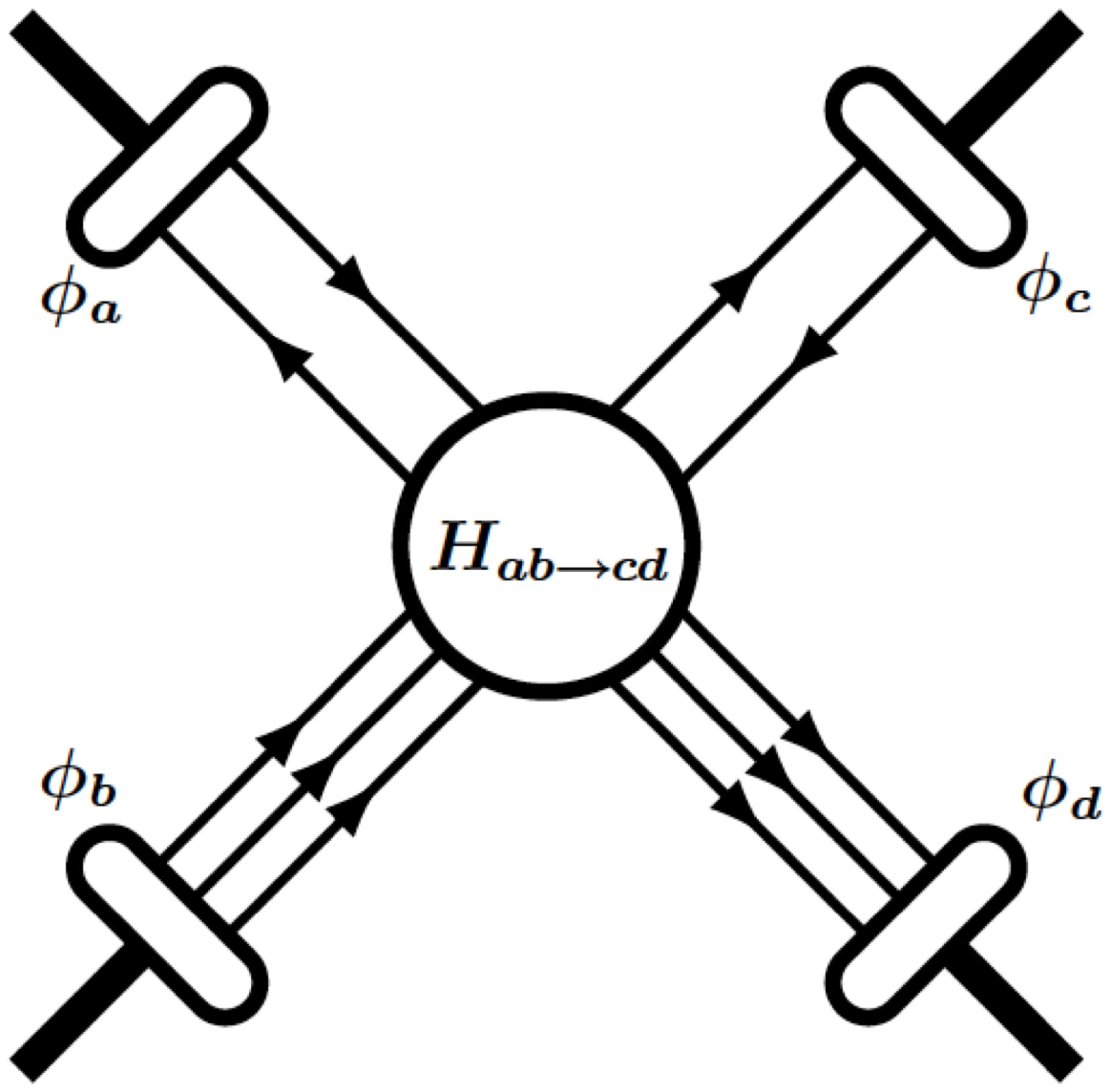}
   \end{center}
\vspace{-0.20cm}
\caption{Exclusive process $a+b \to c+d$ \cite{kks2013}.}
\label{fig:exclusive-ab-cd}
\vspace{-0.5cm}
\end{wrapfigure}

The cross section for a large-angle exclusive scattering
$a+b \to c+d$ is generally expressed in terms of the partonic 
scattering amplitude $H_{ab \to cd}$ and light-cone distribution 
amplitudes $\phi_h  \ (h=a,\,b,\,c,\,d)$,
where the subscript indicates the hadron $h$
as schematically shown in Fig. \ref{fig:exclusive-ab-cd}.
Using the Mandelstam variables $s$ and $t$ defined by
the momenta $p_h$ as
$s = (p_a + p_b)^2$ and $t = (p_a - p_c)^2$, we write 
the cross section as
$ d\sigma_{ab \to cd} / dt \simeq
  \overline{\sum}_{pol} \, | M_{ab \to cd} |^2 / (16 \pi s^2) $
where the matrix element $M_{ab \to cd}$ is expressed as
\cite{exclusive-theory}
\begin{align}
M_{ab \to cd} = & \int [dx_a] \, [dx_b] \, [dx_c] \, [dx_d]  \,
    \phi_c ([x_c]) \, \phi_d ([x_d]) 
\nonumber \\[-0.1cm]
& \ \ \ 
\times 
H_{ab \to cd} ([x_a],[x_b],[x_c],[x_d],Q^2) \, 
      \phi_a ([x_a]) \, \phi_b ([x_b]) .
\label{eqn:mab-cd}
\end{align}
Here, $[x_h]$ indicates a set of the light-cone momentum 
fractions, $x_i=p_i^+/p_h^+$ with $i$-th parton momentum
$p_i$, for partons in the hadron $h$.

There are two factors in the matrix element. One is the parton 
distribution amplitude (PDAs), which reflects a nonperturbative aspect
of hadron, and the other is the partonic scattering amplitude,
which should be calculated in perturbative QCD.
The PDAs have been studied especially for the pion, and there are
theoretical proposals for its functional types, such as
the asymptotic form and the Chernyak-Zhitnitsky form.
They are tested in comparison with measurements of Belle and BaBar
collaborations. Much more studies are needed for other hadron PDAs.

\begin{wrapfigure}[11]{r}{0.46\textwidth}
   \vspace{0.0cm}
   \begin{center}
     \includegraphics[width=5.3cm]{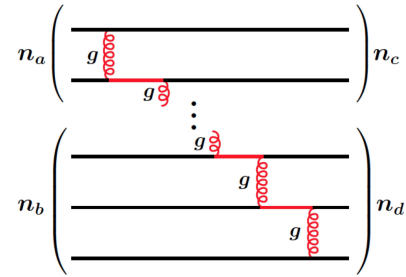}
   \end{center}
\vspace{-0.25cm}
\caption{Typical hard-gluon exchange process \cite{kks2013}.}
\label{fig:hard-glun-exchange}
\vspace{-0.5cm}
\end{wrapfigure}

A typical process is shown in Fig. \ref{fig:hard-glun-exchange}
for the partonic scattering amplitude $H_{ab \to cd}$.
However, depending on the exchanged-gluon location, there are 
many processes which contribute to $H_{ab \to cd}$, and 
an automatic code is needed for such an evaluation.
As shown in Fig. \ref{fig:hard-glun-exchange}, we understand
that the hard exclusive process occurs due to hard gluon 
exchanges to share large momenta among constituents,
which should stick together to form a hadron in an exclusive reaction. 
Then, we obtain so called constituent-counting rule 
by assigning hard momentum factors for the internal quarks,
gluons, and external quarks in  Fig. \ref{fig:hard-glun-exchange}:
\begin{align}
\frac{d\sigma_{ab \to cd}}{dt} = \frac{1}{s^{\, n-2}} \, f_{ab \to cd}(t/s),
\label{eqn:cross-counting}
\end{align}
where $n$ is the total number of constituents involved in the reaction,
$n = n_a+n_b+n_c+n_d$, and $f(t/s)$ is function depends on
the scattering angle.
The cross section scales as $1/s^{\, n-2}$ in the high-energy limit
with the number of constituents, which is called
the constituent-counting rule. 
This rule has been tested by experiments.
Various two-body hadron reactions indicated such a scaling behavior 
in BNL-AGS experiments \cite {counting-exp},
then it is confirmed in photoproduction reactions
such as $\gamma + p \to \pi^+ +n$,
and it is valid even for light nuclei \cite{counting-exp-nuclei}.
We propose to use this idea for the studies of exotic hadrons
\cite{kks2013,cks2016,gpd-gda}. 

\section{Constituent-Counting Rule for Exotic Hadron Candidate}
\label{lambda-1405}

We apply the constituent-counting rule for $\Lambda (1405)$
which has been considered as an exotic hadron for a long time.
Although most low-mass baryons in the 1-2 GeV range can be interpreted
by simple constituent-quark models, the $\Lambda (1405)$ mass is 
anomalously light in the spin-parity $(1/2)^-$ states.
The mass of the ground $\Lambda$ is larger the nucleon mass
($M_\Lambda-M_N \simeq 180$ MeV), whereas the $\Lambda (1405)$ mass
is smaller than the one for a p-wave excitation of the nucleon
($M_{\Lambda (1405)}-M_{N (1535)} \simeq - 130$ MeV)
although $\Lambda (1405)$ contains a heavier-strange quark.
Therefore, $\Lambda (1405)$ is considered as an exotic hadron,
namely a five-quark (pentaquark) hadron or a $\bar K N$ molecule.
If $\Lambda (1405)$ is such a state, its configuration should be 
determined by the scaling behavior in high-energy exclusive 
$\Lambda (1405)$ production processes.

\subsection{$\pi^- + p \to K^0 + \Lambda (1405)$}
\label{hadron-lambda}

\begin{figure}[h!]
\vspace{-0.50cm}
\begin{minipage}{0.48\textwidth}
   \begin{center}
     \includegraphics[width=6.0cm]{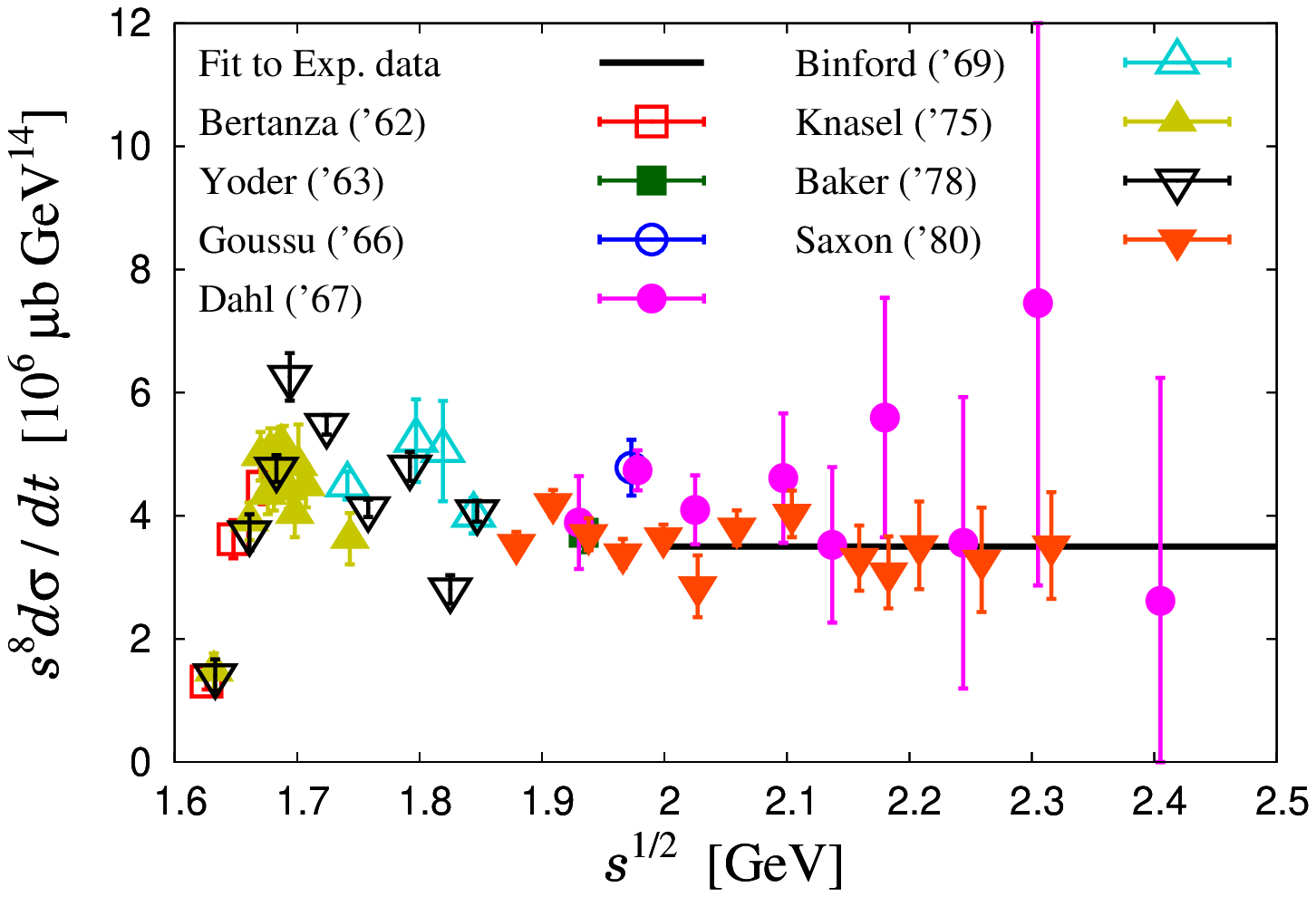}
   \end{center}
\vspace{-0.3cm}
\caption{Scaling in $\pi^- + p \to K^0 + \Lambda$
 \cite{kks2013}.}
\label{fig:pi-lambda}
\end{minipage}
\hspace{0.5cm}
\begin{minipage}{0.48\textwidth}
   \begin{center}
     \includegraphics[width=6.0cm]{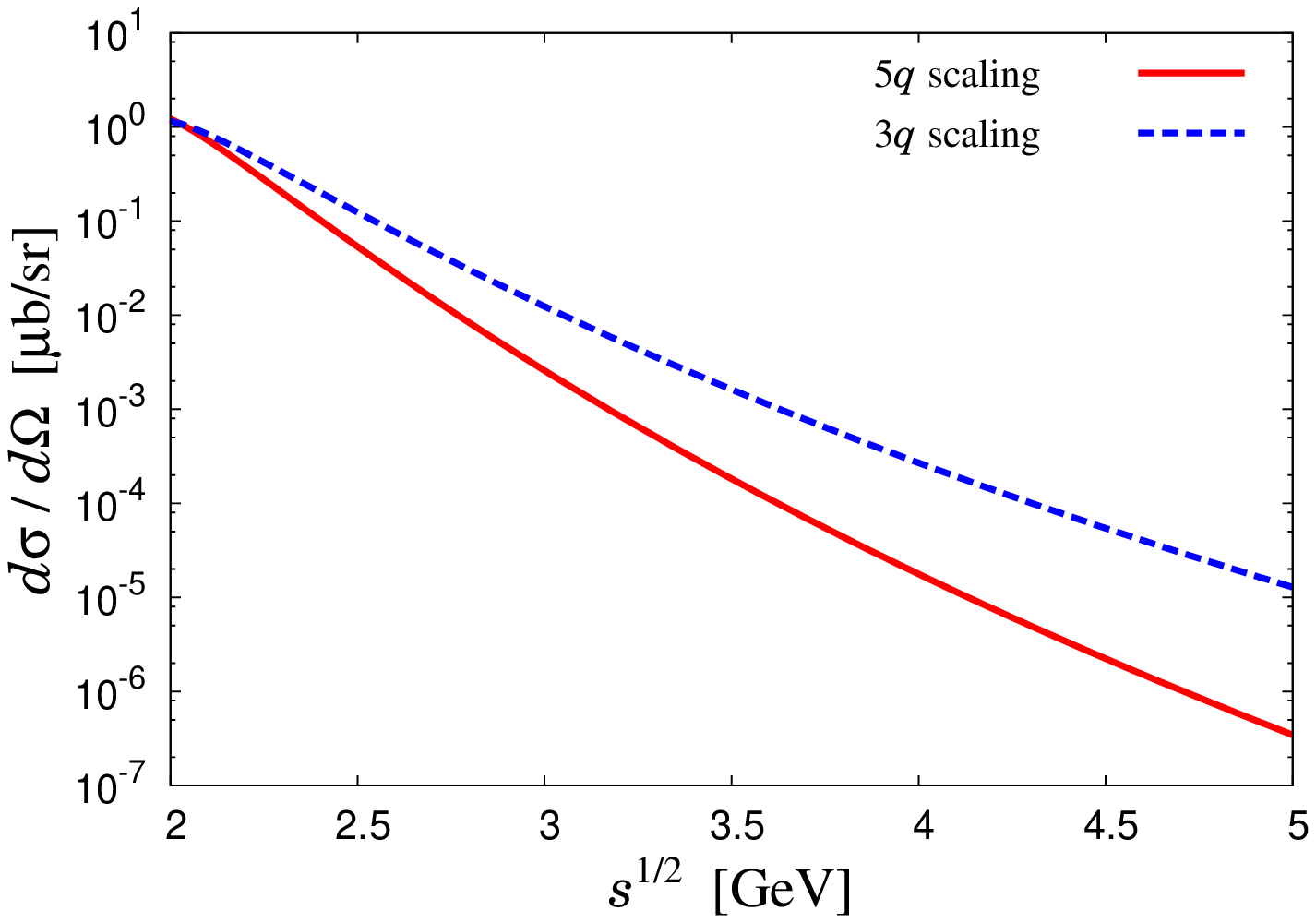}
   \end{center}
\vspace{-0.3cm}
\caption{Scaling in $\pi^- + p \to K^0 + \Lambda (1405)$
 \cite{kks2013}.}
\label{fig:pi-lambda-1405}
\end{minipage} 
\end{figure}

We first discuss pion-induced exclusive $\Lambda$ and $\Lambda (1405)$ 
productions $\pi^- + p \to K^0 + Y$ ($Y=\Lambda,\ \Lambda (1405)$).
Before analyzing $\Lambda (1405)$, we had better test the counting rule
in the ground $\Lambda$ which is considered as an ordinary baryon.
There are many measurement for $\pi^- + p \to K^0 + \Lambda$
and we obtain the cross sections at $\theta_{cm} = 90^\circ$
for studying large-momentum-transfer reactions.
In Fig. \ref{fig:pi-lambda}, the cross sections
multiplied by the scaling factor $s^{8}$ are shown to find whether
it scales by the assignment $n_p=n_\Lambda=3$ and $n_{\pi^-} = n_{K^0}=2$.
In fact, if the factor is determined by the data, we obtain
$n= 10.1 \pm 0.6$, which suggests that the ground $\Lambda$
is consistent with the three-quark picture. 

On the other hand, there is only one data point at $\sqrt{s} = 2.02$ GeV
for $\Lambda (1405)$.
It is not obvious where the scaling region starts for the $\Lambda (1405)$
production; however, the transition occurs at $\sqrt{s} \simeq 2.5$ GeV
according to the experimental data for $\gamma + p \to \pi^+ + n$
\cite{counting-exp}.
Unless the full evaluation of Eq. (\ref{eqn:mab-cd}) becomes
possible, we cannot predict the transition point theoretically.
So, simply assuming that the scaling description works 
at $\sqrt{s} \ge 2.02$ GeV, we draw two cross section curves
in Fig. \ref{fig:pi-lambda-1405} by taking
$\Lambda (1405)$ as a three-quark or five-quark state
for a rough estimate on future measurements.
In fact, an experimental proposal is under consideration at J-PARC
by using the high-momentum beamline \cite{j-parc-exp}. 
If the cross sections will be measured at least at two different 
energies, they should provide information on the number of 
constituents in $\Lambda (1405)$ in a different way from 
low-energy observables on masses and decay widths.

\vfill\eject

\subsection{$\gamma + p \to K^+ + \Lambda (1405)$}
\label{hadron-lambda}

\begin{figure}[h!]
\vspace{-0.5cm}
\begin{minipage}{0.48\textwidth}
   \begin{center}
     \includegraphics[width=6.0cm]{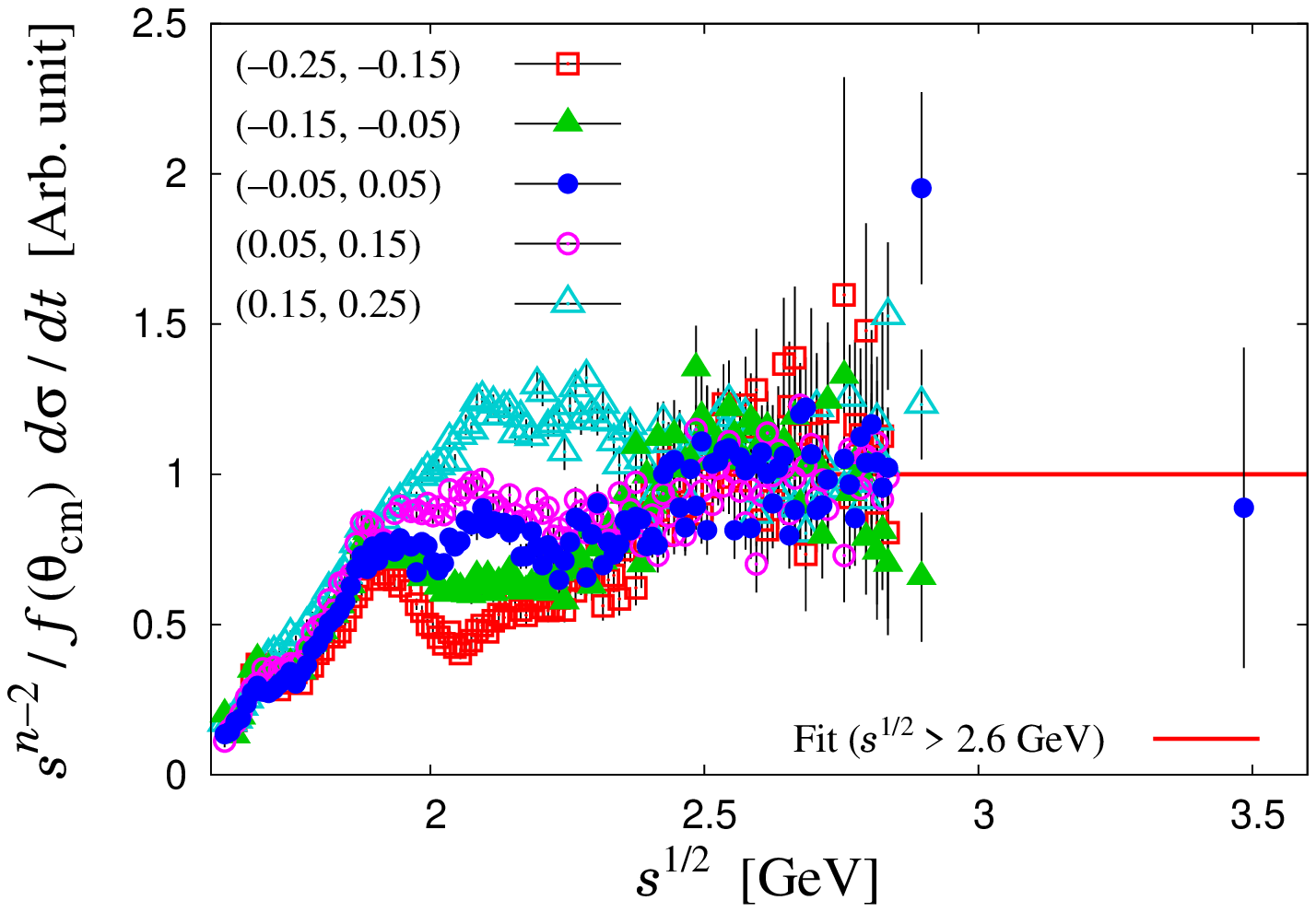}
   \end{center}
\vspace{-0.45cm}
\caption{Scaling in $\gamma + p \to K^+ + \Lambda$
 \cite{cks2016}.}
\label{fig:gamma-lambda}
\end{minipage}
\hspace{0.5cm}
\begin{minipage}{0.48\textwidth}
   \begin{center}
     \includegraphics[width=6.0cm]{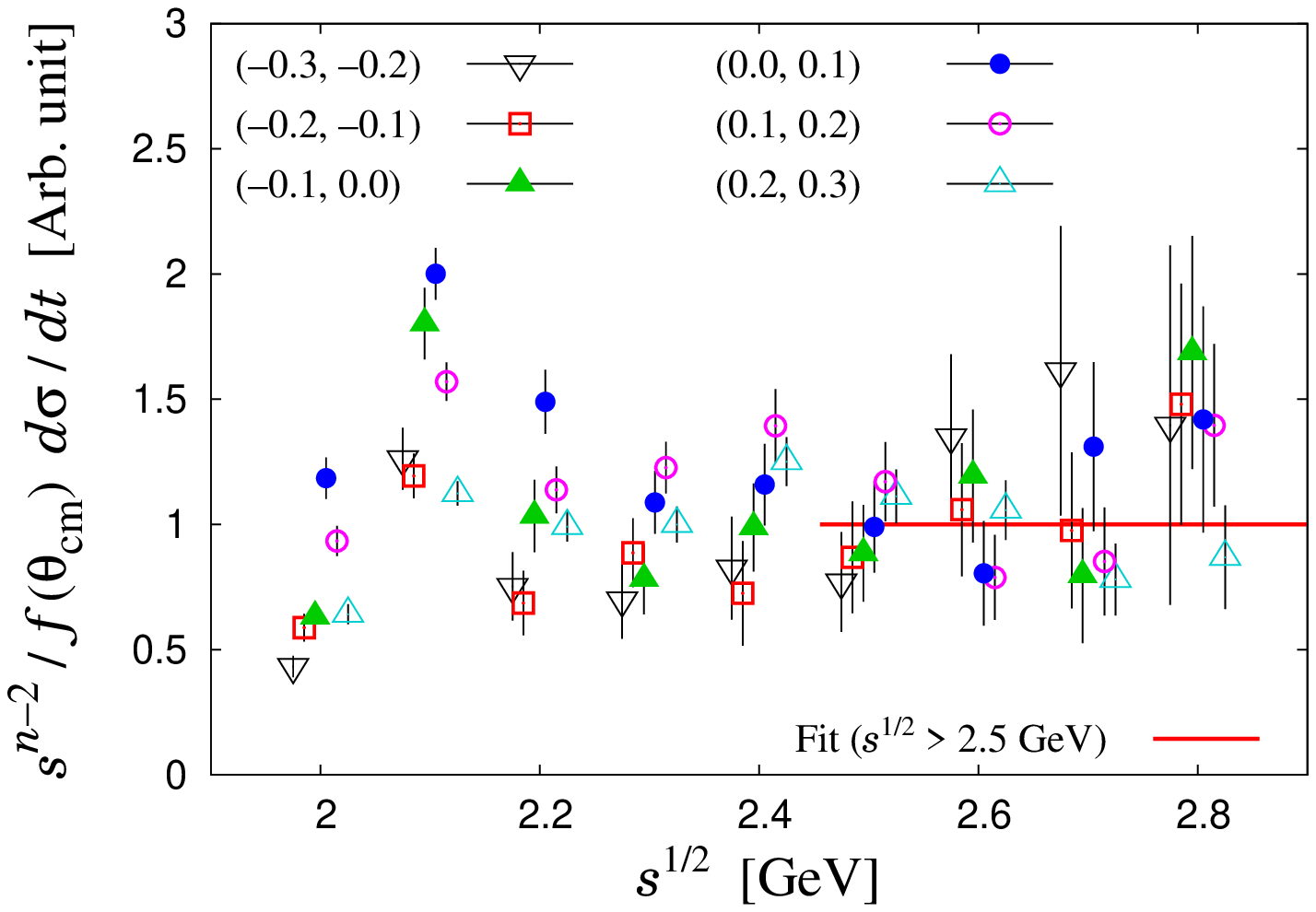}
   \end{center}
\vspace{-0.45cm}
\caption{Scaling in $\gamma + p \to K^+ + \Lambda (1405)$
 \cite{cks2016}.}
\label{fig:gamma-lambda-1405}
\end{minipage} 
\vspace{-0.5cm}
\end{figure}

We discuss photoproductions of hyperons including $\Lambda (1405)$.
We determined the factor $n$ by fitting the data, which are shown in 
Fig. \ref{fig:gamma-lambda} for the reaction $\gamma+p \to K^+ + \Lambda$.
We take measurements in scattering-angle bins around 
$\theta _{\rm cm} = 90^{\circ}$.
In Fig. \ref{fig:gamma-lambda}, five data sets are shown for the bins 
from $-0.25 < \cos \theta _{\rm cm} < -0.15$ to 
$0.15 < \cos \theta _{\rm cm} < 0.25$ with $0.10$ step each.
Using the data with $s \ge s_{min}$ with 
$\sqrt{s_{min}}=2.3$, 2.4, $\cdots$, 2.7 GeV,
we determined the scaling factor $n$.
In Fig. \ref{fig:gamma-lambda}, the cross sections with $n=10.0$ are
shown from the analysis of $\sqrt{s} \ge \sqrt{s_{min}}=2.6$ GeV.
The data at $\sqrt{s}=3.5$ GeV was taken from a SLAC experiment,
and low-energy data ware from the JLab-CLAS experiment.
The measurements tend to indicate scaling behavior at high energies;
however, it should become more clearly if accurate measurements 
will be taken in the intermediate-energy region
(2.8 GeV $\le \sqrt{s} \le 3.5$ GeV) at the updated 12 GeV facility
of JLab.
Next, $\Lambda (1405)$ production data are shown in 
Fig. \ref{fig:gamma-lambda-1405}, where $n=10.6$ in the analysis
range of $\sqrt{s} \ge 2.5$ GeV. The data have large errors and
there is no measurement at 3.5 GeV, so that the scaling behavior is not
very clear. Obviously, much accurate and higher-energy data are
needed for the scaling analysis.

\begin{wrapfigure}[11]{r}{0.46\textwidth}
   \vspace{-0.2cm}
   \begin{center}
     \includegraphics[width=5.7cm]{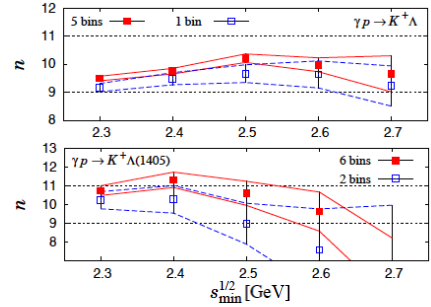}
   \end{center}
\vspace{-0.30cm}
\caption{Energy dependence of factor $n$ \cite{cks2016}.}
\label{fig:cross-section}
\vspace{-0.5cm}
\end{wrapfigure}

We investigated the minimum-energy ($\sqrt{s_{min}}$) dependence 
on the scaling factor $n$ and the results are shown in 
Fig. \ref{fig:cross-section}. 
By varying the value of $\sqrt{s_{min}}$, the factor $n$ is 
determined for $\Lambda$ and $\Lambda (1405)$. 
If the hyperon is a three-quark or five-quark state, $n$
should be 9 or 11, respectively. The results show interesting phenomena.
The ground $\Lambda$ is roughly consistent with the three-quark picture,
whereas $\Lambda (1405)$ looks like five-quark state at low energies
(2.3 GeV $\le \sqrt{s_{min}} \le 2.5$ GeV) and it becomes three-quark one 
at high energies ($\sqrt{s_{min}} \ge 2.6$ GeV). Therefore, it seems that
$\Lambda (1405)$ looks like five-quark (or $\bar KN$ molecule) 
at relatively low energies but it is more like three-quark state
at high energies. It is an interesting finding; however, 
we cannot draw a solid conclusion at this stage because 
the errors of the $n$ values are too large.

\section{Exotic Hadron Studies by Three-Dimensional Structure Functions}
\label{3d}

Internal structure of exotic hadron candidates can be also investigated
by three-dimensional structure functions \cite{gpd-gda}, 
namely generalized parton distributions (GPDs), 
transverse-momentum-dependent parton distributions (TMDs), 
and generalized distribution amplitudes (GDAs) \cite{gpd-gda-summary}. 
The constituent counting rule can be also used as a useful tool
in these investigations for exotic hadrons.
First, the GPDs for the nucleon are defined by off-forward matrix elements
of quark (and gluon) operators as
\begin{align}
 \! \! \! \! \! \! 
 \int \!  \frac{d y^-}{4\pi}e^{i x \bar P^+ y^-} \! \!
 \left< p' \left| 
 \overline{\psi}(-y/2) \gamma^+ \psi(y/2) 
 \right| p \right> \Big |_{y^+ = \vec y_\perp =0}
 \! \! = \!
 \frac{1}{2  \bar P^+} \, \overline{u} (p') 
 \left [ H_q (x,\xi,t) \gamma^+
    \! \! + \! E_q (x,\xi,t)  \frac{i \sigma^{+ \alpha} \Delta_\alpha}{2 \, M}
 \right ] u (p) ,
\label{eqn:gpd-n}
\end{align}
where $H_q (x,\xi,t)$ and $E_q (x,\xi,t)$ are the GPDs.
The GPDs are experimentally measured by the virtual Compton process.
The kinematical variables $x$, $\xi$, and $t$ are defined as follows.
We define momenta $\bar P$, $\bar q$, and $\Delta$ by
the nucleon and photon momenta as
$\bar P = (p+p')/2$,
$\bar q = (q+q')/2$, 
$\Delta = p'-p = q-q'$, and the momentum-transfer-squared quantities 
are given by $Q^2 = -q^2$, $\bar Q^2 = - \bar q^2$, and $t = \Delta^2$.
Then, the generalized scaling variable $x$ and a skewdness parameter $\xi$ 
are defined by $x   = Q^2/(2p \cdot q)$ and
$\xi = \bar Q^2 /(2 \bar P \cdot \bar q)$.

\begin{wrapfigure}[12]{r}{0.42\textwidth}
   \vspace{-0.1cm}
   \begin{center}
     \includegraphics[width=5.1cm]{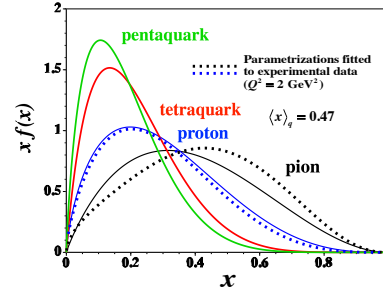}
   \end{center}
\vspace{-0.30cm}
\caption{Valence-quark distributions for pion, proton,
tetraquark, and pentaquark \cite{gpd-gda}.}
\label{fig:exotic-pdf}
\vspace{-0.5cm}
\end{wrapfigure}

The GPDs contain information on both PDFs ($f_n (x)$)
and transverse form factors ($F_n^h (t, x) $),
and they are typically expressed by their multiplication:
\begin{align}
H_q^h (x,\xi=0,t)= f_n (x) \, F_n^h (t, x) ,
\label{eqn:gpd-paramet1}
\end{align}
where $n$ indicates the number of valence quarks.
As for the longitudinal PDFs, a simple $x$-dependent form
$ f_n (x) = C_n \, x^{\alpha_n} \, (1-x)^{\beta_n} $
is often used. 
For a valence-quark distribution, 
the parameters $C_n$, $\alpha_n$, and $\beta_n$ are constrained by
the valence-quark number
$ \int_0^1 dx \, f_n (x) = n$, and the quark momentum
$ \int_0^1 dx \, x \, f_n (x) = \langle x \rangle_q$.
The parameter $\beta_n$ controls the behavior in the elastic limit
$x \to 1$, and it is determined by the constituent counting rule 
as $\beta_n = 2n -3+2\Delta S_z$ with the spin factor 
$\Delta S_z=|S_z^q-S_z^h|$. By these three conditions, we 
theoretically obtain the solid curves for tetraquark and
pentaquark hadrons as well as the pion and proton.
As for the typical PDFs of the pion and proton extracted from
experimental measurements, the dotted and dashed curves are shown
at $Q^2=2$ GeV$^2$ for comparison.
They roughly agree with the simple theoretical curves,
so that it is interesting to measure the PDFs for exotic hadrons
shown in Fig. \ref{fig:exotic-pdf}.
In addition, the transverse form factor $F_n^h (t, x)$ also
contains information on exotic nature \cite{gpd-gda}.

\begin{figure}[h!]
\vspace{-0.1cm}
\hspace{1.5cm}
\begin{minipage}{0.48\textwidth}
     \includegraphics[width=5.0cm]{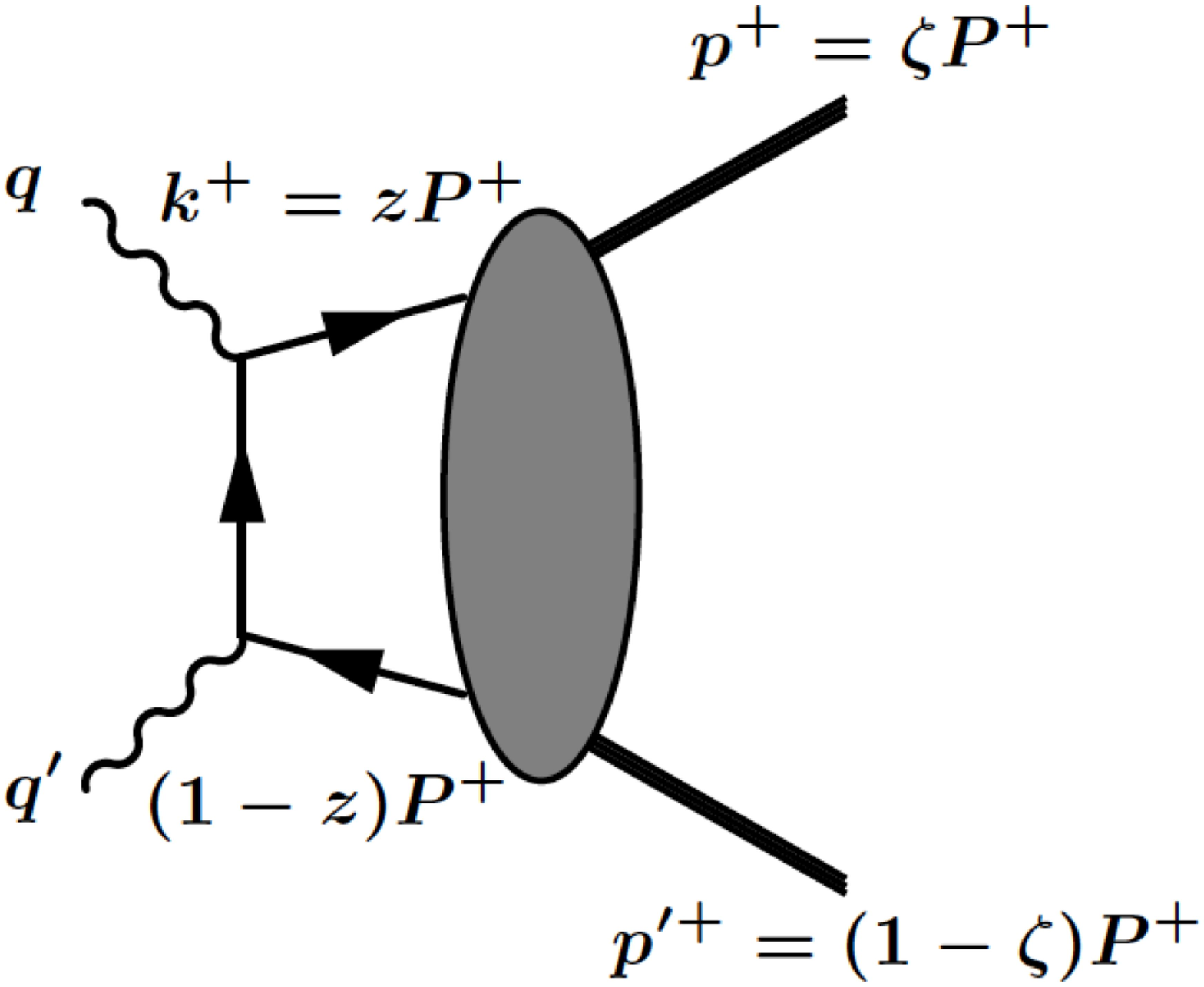}
\vspace{-0.0cm}
\caption{$\gamma\gamma \to h \bar h$ and GDAs\cite{gpd-gda}.}
\label{fig:dga-1}
\end{minipage}
\hspace{-1.0cm}
\begin{minipage}{0.50\textwidth}
\vspace{-0.3cm}
   \begin{center}
     \includegraphics[width=6.0cm]{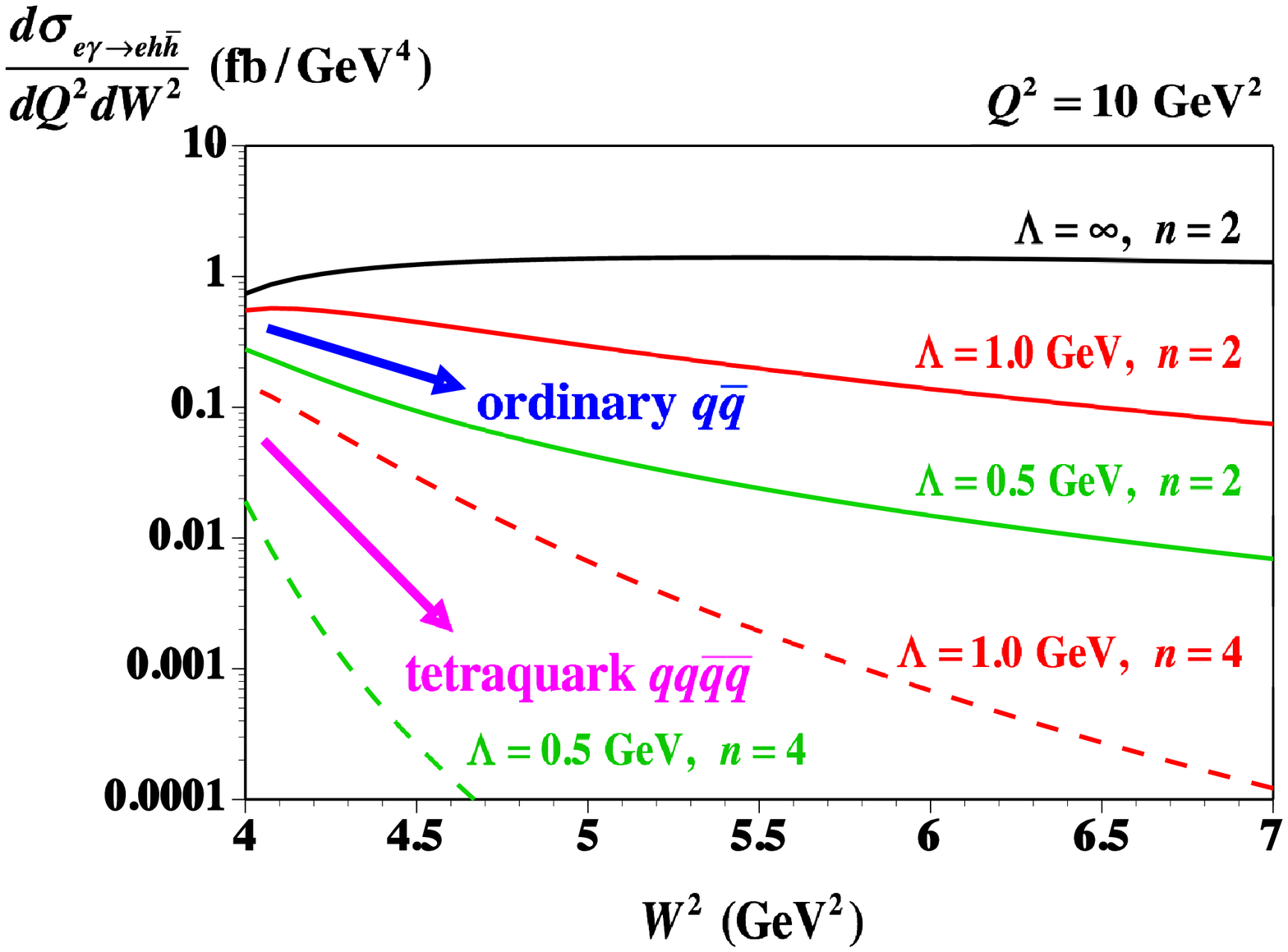}
   \end{center}
\vspace{-0.16cm}
\caption{Cross section for $h\bar h$-pair production \cite{gpd-gda}.}
\label{fig:two-photon}
\end{minipage} 
\vspace{-0.8cm}
\end{figure}

Since there is no stable exotic hadron candidate which can be
used for a fixed-target experiment, these interesting properties
of exotic hadrons in the GPDs may not be observed directly.
There is a possibility to investigate them in the transition GPDs
such as $p \to \Lambda (1405)$. On the other hand, three-dimensional
structure functions can be observed even for exotic hadron candidates,
namely unstable hadrons, in two-photon processes 
to produce an exotic-hadron pair $\gamma\gamma^* \to h \bar h$
as shown in Fig. \ref{fig:dga-1}.
The factorization of the process, by the hard process 
of photon interactions with quarks and the soft part with the GDAs,
works in the kinematical region $Q^2 \gg W^2, \, \Lambda^2$.
The GPDs are mainly measured in the deeply virtual Compton scattering
(DVCS) $\gamma^* h \to \gamma h$, so that the two photon process is
the $s$-$t$ crossed one to the DVCS. The $s$-$t$ crossed three-dimensional
structure functions to the GPDs are called GDAs.
They should be able to provide an important information on
internal structure of exotic hadron candidates.
The quark GDAs are defined by the same lightcone operator, as the one 
in the GPD definition, between the vacuum and the hadron pair $h\bar h$:
\begin{align}
\Phi_q^{h\bar h} (z,\zeta,s) 
= \int \frac{d y^-}{2\pi}\, e^{i (2z-1)\, P^+ y^-}
   \langle \, h(p) \, \bar h(p') \, | \, 
 \overline{\psi}(-y/2) \gamma^+ \psi(y/2) 
  \, | \, 0 \rangle \Big |_{y^+=\vec y_\perp =0} \, ,
\end{align}
The gluon GDA is defined in the similar way.
The $s$-$t$ crossing indicates to move the final state 
$\bar h$ ($p'$) to the initial $h$ ($p$), so that both quantities
are related to each other by
\begin{align}
\Phi_q^{h\bar h} (z,\zeta,W^2) 
\longleftrightarrow
H_q^h \left ( x=\frac{1-2z}{1-2\zeta},
            \xi=\frac{1}{1-2\zeta}, t=W^2 \right ) .
\label{eqn:gda-gpd-relation}
\end{align}
From this relation, we notice that the kinematical regions,
$ 0 \le |x| < \infty$, $ 0 \le |\xi| < \infty$, 
$ |x| \le |\xi|$,      $ t \ge 0 $, 
contain unphysical regions of the GDAs.
Therefore, the GPD studies are not necessarily connected directly 
for the physical GPDs and vice versa.
In the same way with the GPDs, a simple function
\begin{align}
\Phi_q^{h\bar h (I=0)} (z,\zeta,W^2) 
= N_{h(q)} \, z^\alpha (1-z)^\beta (2z-1) \, \zeta (1-\zeta) \, F_{h(q)} (W^2) ,
\label{eqn:gda-paramet}
\end{align}
is often used for the GDAs.
Here, $F_{h(q)} (W^2)$ indicates a form factor of the quark part
of the energy-momentum tensor, and it could be taken as
$ F_{h(q)} (W^2) 
= 1 / [ 1 + (W^2-4 m_h^2)/\Lambda^2 ]^{n-1} $
according to the constituent counting rule.
The factor $n$ is $n=2$ for ordinary $q\bar q$ mesons and $n=4$ 
for tetraquark hadrons.

We estimated $W^2$ dependence of the cross section 
$e \gamma \to e' h \bar h$ by using the simple GDAs 
with the form factor suggested by the constituent counting rule,
and the results are shown in Fig. \ref{fig:two-photon}
for $h=f_0 (980)$ or $a_0 (980)$.
Depending on the constituent number $n$, namely 
$n=2$ (ordinary meson, $q\bar q$) or 
$n=4$ (tetraquark type, $qq\bar q \bar q$),
there are large differences between the $n=2$ and $n=4$
curves, so that exotic nature of the hadron $h$ should be
clarified. Furthermore, even if the hadron has $n=4$,
we could distinguish a compact $qq\bar q \bar q$ state
from a diffuse molecular state from the measurement.

\section{Summary}
\label{summary}
\vspace{-0.0cm}

The constituent counting rule was predicted in perturbative QCD,
and it is confirmed by various high-energy exclusive processes.
We proposed to use this rule for finding internal quark 
configurations of exotic hadron candidates. 
Three-dimensional structure functions, such as GPDs and TMDs, 
have been studied recently for clarifying the origin of nucleon spin, 
especially the partonic orbital momentum contributions. 
Here, we proposed that the GPDs and GDAs should be valuable
for probing internal structure of exotic hadron candidates.


\vspace{-0.0cm}
\section*{Acknowledgements}
\vspace{-0.3cm}
This work was partially supported by JSPS KAKENHI Grant Numbers 
JP25105010 and JP15K05061.



\end{document}